\documentclass[3p,times]{elsarticle}
\usepackage{amssymb}
\usepackage{amsxtra}
\usepackage{amsmath}
\usepackage{amsfonts}
\usepackage{CJK}
\usepackage[titletoc]{appendix}
\usepackage{graphicx}
\usepackage{hyperref}
\usepackage{cleveref}
\numberwithin{equation}{section}

\newtheorem{thm}{Theorem}[section]

\newtheorem{prop}[thm]{Proposition}

\newtheorem{ass}[thm]{Assumption}

\newtheorem{de}[thm]{Definition}
\newtheorem{rem}[thm]{Remark}

\newcommand{\eqa}{\begin{eqnarray}}
\newcommand{\eeqa}{\end{eqnarray}}
\newcommand{\beq}{\begin{equation}}
\newcommand{\eeq}{\end{equation}}
\newcommand{\nn}{\nonumber}
\newcommand{\p}{\partial}

\allowdisplaybreaks
\begin{document}
\begin{frontmatter}
\title{Riemann-Hilbert problem associated with the fourth-order dispersive nonlinear Schr\"{o}dinger equation in optics and magnetic mechanics}
\author[BB]{Beibei Hu\corref{cor1}}
\author[BB]{Ling Zhang\corref{cor1}}
\author[BB]{Qinghong Li}
\author[ZN]{Ning Zhang}
%\ead{This research is partially supported by NSF of China (Grant numbers 11372170).}

\cortext[cor1]{Corresponding authors at: School of Mathematics and Finance, Chuzhou University, Anhui 239000, China. \\
 Email addresses: hu\_chzu@shu.edu.cn(B.-B. Hu), originzhang@126.com(L. Zhang)}

\address[BB]{School of Mathematics and Finance, Chuzhou University, Anhui, 239000, China}
%\address[TC]{Department of Mathematics, Shanghai University, Shanghai 200444, China}
\address[ZN]{Department of Basical Courses, Shandong University of Science and Technology, Taian 271019, China}
\pagestyle{plain}
\setcounter{page}{1}
\begin{abstract}
In this paper, we utilize Fokas method to investigate the initial-boundary value problems (IBVPs) of the fourth-order dispersive nonlinear Schr\"{o}dinger (FODNLS) equation on the half-line, which can simulate the nonlinear transmission and interaction of ultrashort pulses in the high-speed optical fiber transmission system, and describe the nonlinear spin excitation phenomenon of one-dimensional Heisenberg ferromagnetic chain with eight poles and dipole interaction. By discussing the eigenfunctions of Lax pair of FODNLS equation and the analysis and symmetry of the scattering matrix, the IBVPs of FODNLS equation is expressed as a matrix Riemann-Hilbert (RH) problem form. Then one can get the potential function solution $u(x,t)$ of the FODNLS equation by solving this matrix RH problem. In addition, we also obtained that some spectral functions admits a key global relationship.
\end{abstract}
\end{frontmatter}

\section{Introduction}
For a long time, finding a method to solve integrable equations has been a very important research topic in theory and application. The display of integrable equations with exact solutions and some special solutions can provide important guarantees for the analysis of its various properties. However, there is no unified method to solve all integrable equations. With the in-depth study of integrable systems by scholars, a series of methods to solve the classic integrable development equation have emerged. For example, inverse scattering method\cite{GGKM1967}, Hirota method\cite{HR1971}, B\"{a}cklund transform\cite{WHD1973}, Darboux transform(DT)\cite{MVB1991} and so on. Among them, the IST method is the main analytical method for the exact solution of nonlinear integrable systems. However, due to the IST method is suitable for the limitations of the initial value conditions at infinity, and it is almost only used to study the pure initial value problem of integrable equations, many real-world phenomena and some studies in the fluctuation process not only need to consider the initial value conditions, but the boundary value conditions also need to be considered. Naturally, people need to replace the initial value problems with the initial-boundary value problems(IBVPs) in the research process.

In 1997, Fokas proposed a unified transformation method from the initial value problem to the IBVPs based on the IST method idea, which is called the Fokas method. This method can be investigated IBVPs of partial differential equation\cite{Fokas1997}, and in the past 22 years, IBVPs of some classical integrable equations to be discussed via the Fokas method. For example, the modified Korteweg-de Vries(MKdV) equation\cite{Fokas2002}, the nonlinear Schr\"{o}dinger(NLS) equation\cite{Fokas2005}, the Kaup-Newell equation\cite{Lenells2008}, the stationary axisymmetric Einstein equations\cite{lenells2011}, the Ablowitz-Ladik system\cite{Xia2017}, the Kundu-Eckhaus equation\cite{Hu2019}, the Hirota equation\cite{CSY2019,HL2020}. In 2012, Lenells extended the Fokas method to the integrable equation with higher-order matrix spectrum, he proposed a more general unified transformation approach to solving IBVPs of integrable model\cite{Lenells2012} and using the unified transformation approach to analyzed IBVPs of Degasperis-Procesi equation\cite{Lenells2013}. After that, more and more individuals began to study the IBVPs of integrable model with higher-order matrix spectral\cite{Monvel2013,Xu2013,Xu2014,Geng2015,Liu2016,Tian2017,Yan2017,Yan2019,ZQZ2017}. The authors have also done a slice of works on the application of the Fokas method to an integrable equation with higher-order matrix Lax pairs\cite{Hu1,Hu2,Hu3}.

In this paper, our work is related to the fourth-order dispersive nonlinear Schr\"{o}dinger(FODNLS) equation\cite{PK1992,CA2013} expressed as:
\eqa
iu_{t}+\alpha_1u_{xx}+\alpha_2u|u|^2+\frac{\varepsilon^2}{12}(\alpha_3u_{xxxx}+\alpha_4|u|^2u_{xx}
+\alpha_5u^2\overline{u}_{xx}+\alpha_6u_{x}^2\overline{u}+\alpha_7u|u_{x}|^2+\alpha_8|u|^4u)=0,
\label{1.1}\eeqa
where $u$ represents the amplitude of the slowly varying envelope of the wave, $x$ and $t$ are the normalized space and time variables, $\varepsilon^2$ is a dimensionless small parameter representing the high-order linear and nonlinear strength, and $\alpha_j(j=1,2,\ldots,8)$ is the real parameter.
The Eq.\eqref{1.1} is mainly derived from fiber optics and magnetism. On the one hand, in optics, Eq.\eqref{1.1} can simulate the nonlinear propagation and interaction of ultrashort pulses in high-speed fiber-optic transmission systems\cite{AF2009}. On the other hand, in magnetic mechanics, Eq.\eqref{1.1} can be used to describe the nonlinear spin excitation of a one-dimensional Heisenberg ferromagnetic chain with octuple and dipole interactions\cite{DM2001}. In particular, when the parameter value is $\alpha_1=\alpha_3=1, \alpha_2=\alpha_5=2, \alpha_4=8, \alpha_6=\alpha_8=6, \alpha_7=4$, and let $\gamma=\frac{\varepsilon^2}{12}$ the Eq.\eqref{1.1} becomes to
\eqa
iu_{t}+u_{xx}+2u|u|^2+\gamma(u_{xxxx}+8|u|^2u_{xx}+2u^2\overline{u}_{xx}+6u_{x}^2\overline{u}+4u|u_{x}|^2+6|u|^4u)=0,
\label{1.2}\eeqa
which is an integrable model, and many properties have been widely studied, such as, the Lax pair, the infinite conservation laws\cite{ZHQ2009}, the breather solution, and the higher-order rogue wave solution based on the DT method\cite{WXL2012,WLH2013,YB2013}, the multi-soliton solutions by using Riemann-Hilbert(RH) approach\cite{LWH2019}, the dark and bright solitary waves and rogue wave solution by using phase plane analysis method\cite{LM2020}, the bilinear form and the N-soliton solution via the Hirota approach\cite{LRX2013,LRX2014}. However, as far as we know, the FODNLS \eqref{1.2} on the hale-line has not been studied, and in the following work, we utilize Fokas method to discuss the IBVPs of the FODNLS equation\eqref{1.2} on the half-line domain $\Omega=\{(x,t): 0<z<\infty,0<t<T\}$.

The paper is organized as follows. In section 2, one can introducing eigenfunction to spectral analysis of the Lax pair. In section 3, a slice of key functions $y(\zeta), z(\zeta), Y(\zeta), Z(\zeta)$ are further discussed. In section 4, an important theorem is proposed. And the last section is devoted to conclusions.

\section{The spectral analysis}

Base on Ablowitz-Kaup-Newell-Segur scheme, the Lax pair of Eq.\eqref{1.2} is expressed as\cite{ZHQ2009,WXL2012,WLH2013,YB2013,LWH2019}
\begin{subequations}
\begin{align}
&\Psi_{x}=(-i\zeta\Lambda+P)\Psi,\label{2.1a} \\
&\Psi_{t}=[(8i\gamma\zeta^{4}-2i\zeta^2)\Lambda-8\gamma\zeta^3P-4i\gamma\zeta^2A_1-2\zeta A_2+iA_3]\Psi,\label{2.1b}
\end{align}
\end{subequations}
where $\zeta$ is a complex spectral parameter, $\Psi=(\Psi_1,\Psi_2)^T$ is the vector eigenfunction, the $2\times2$ matrices $\Lambda=diag\{1,-1\}$, and $P, A, B$ and $C$ are defined by
\eqa \begin{array}{l}
P=\left(\begin{array}{cc}
0 & u \\
-\overline{u} & 0
\end{array}\right),
A_1=\left(\begin{array}{cc}
|u|^2 & u_x \\
\overline{u}_x & -|u|^2
\end{array}\right),
A_2=\left(\begin{array}{cc}
\gamma(u\overline{u}_x-\overline{u}u_x) & -\gamma u_{xx}-(2\gamma|u|^2+1)u \\
 -\gamma \overline{u}_{xx}-(2\gamma|u|^2+1)\overline{u} & -\gamma(u\overline{u}_x-\overline{u}u_x)
\end{array}\right),\\
A_3=\left(\begin{array}{cc}
\gamma(3|u|^4-|u_x|^2+\overline{u}u_{xx}+u\overline{u}_{xx})+|u|^2 & \gamma u_{xxx}+(6\gamma|u|^2+1)u_x \\
 \gamma \overline{u}_{xxx}+(6\gamma|u|^2+1)\overline{u}_x & -\gamma(3|u|^4-|u_x|^2+\overline{u}u_{xx}+u\overline{u}_{xx})+|u|^2
\end{array}\right).
\end{array} \label{2.2}\eeqa

\subsection{The exact one-form}

The Lax pair equations \eqref{2.1a}-\eqref{2.1b} are rewritten as follows
\begin{subequations}
\begin{align}
&\Psi_{x}+i\zeta\Lambda\Psi=P(x,t,\zeta)\Psi,\label{2.3a} \\
&\Psi_{t}-(8i\gamma\zeta^{4}-2i\zeta^2)\Lambda\Psi=R(x,t,\zeta)\Psi,\label{2.3b}
\end{align}
\end{subequations}
where
\eqa && R(x,t,\zeta)=-8\gamma\zeta^3P-4i\gamma\zeta^2A_1-2\zeta A_2+iA_3
\nn\\&&\quad=-8\gamma\zeta^3P+4i\gamma\zeta^2(P^2+P_x)\Lambda+2\gamma\zeta(PP_x-P_xP-P_{xx}+2P^3)\Lambda-2\zeta P\Lambda
\nn\\&&\qquad+i\gamma(3P^4+P_x^2-P_{xx}P-PP_{xx}-P_{xxx}+6P^2P_x-P_x)\Lambda-iP^2.\nn\eeqa

For the convenience of later calculation, we record $\theta=(8\gamma\zeta^4-2\zeta^2)$ and introduce the following function transformation
\eqa \Psi(x,t,\zeta)=G(x,t,\zeta) e^{i[(8\gamma\zeta^4-2\zeta^2) t-\zeta x]\Lambda},0<x<\infty, 0<t<T. \label{2.4}\eeqa
Then, we get
\begin{subequations}
\begin{align}
&G_x+i\zeta[\Lambda,G]=PG,\label{2.5a}\\
&G_t-i(8\gamma\zeta^4-2\zeta^2)[\Lambda,G]=RG,\label{2.5b}
\end{align}
\end{subequations}
which can be expressed as the following full differential
\eqa d(e^{i[\zeta x-(8\gamma\zeta^4-2\zeta^2) t]\hat{\Lambda}}G(x,t,\zeta))=F(x,t,\zeta),\label{2.6}\eeqa
where exact one-form $F(x,t,\zeta)$ is
\eqa F(x,t,\zeta)=e^{i[\zeta x-(8\gamma\zeta^4-2\zeta^2) t]\hat{\Lambda}}(P(x,t,\zeta)dx+R(x,t,\zeta)dt)G(x,t,\zeta),\label{2.7}\eeqa
and $\hat{\Lambda}$ represents a matrix operator acting on a second order matrix $\Lambda$, i.e. $\hat{\Lambda}P=[\Lambda,P]$ and $e^{\hat{\Lambda}}P=e^{\Lambda}Pe^{-\Lambda}$.

\subsection{ The analytic and bounded eigenfunctions $G_j(x,t,\zeta)'s$}

We assume that $u(x,t)\in \mathcal{S}$ with $(x,t)\in\Omega=\{(x,t): 0<z<\infty,0<t<T\}$, and use the integral equation containing the exact one-form to define eigenfunctions $\{G_j(x,t,\zeta)\}_1^3$ of Eq.\eqref{2.5a}-\eqref{2.5b} as follows
\eqa G_j(x,t,\zeta)=\mathrm{I}+\int_{(x_j,t_j)}^{(x,t)}e^{i[(8\gamma\zeta^4-2\zeta^2) t-\zeta x]\hat{\Lambda}}F(\xi,\tau,\zeta),  \label{2.18}\eeqa
where the integration path is $(x_j,t_j)\rightarrow(x,t)$ which is a directed smooth curve. It follows from the closed of the exact one-form that the integral of Eq.\eqref{1.2} is independent of the integration path. Therefore, one can choose three integral curve are all parallel to the axis shown in Figure 1.

\begin{figure}
\centering
\includegraphics[width=4.0in,height=1.0in]{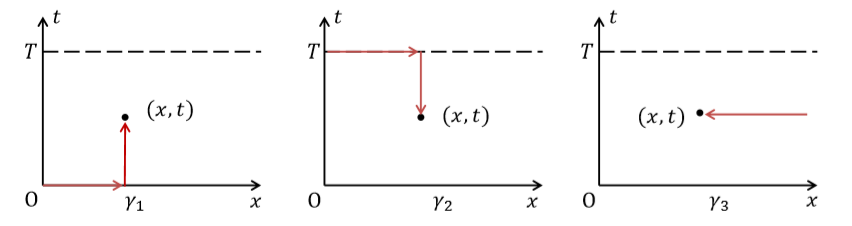}
\caption{The smooth curve $\gamma_1,\gamma_2,\gamma_3$ in the $(x,t)$-plan}
\label{fig:graph}
\end{figure}

We might as well take $(x_1,t_1)=(0,0), (x_2,t_2)=(0,T)$, and $(x_3,t_3)=(\infty,t)$, then we have
\begin{subequations}
\begin{align}
&G_1(x,t,\zeta)=\mathrm{I}+\int_{0}^{x}e^{-i\zeta(x-\xi)\hat{\Lambda}}(PG_1)(\xi,t,\zeta)d\xi\nn\\
&\qquad\qquad\qquad
+e^{-i\zeta x\hat{\Lambda}}\int_{0}^{t}e^{i(8\gamma\zeta^4-2\zeta^2) (t-\tau)\hat{\Lambda}}(R_1G_1)(0,\tau,\Lambda)d\tau,\label{2.19a}\\
&G_2(x,t,\zeta)=\mathrm{I}+\int_{0}^{x}e^{-i\zeta(x-\xi)\hat{\Lambda}}(PG_2)(\xi,t,\zeta)d\xi\nn\\
&\qquad\qquad\qquad
-e^{-i\zeta x\hat{\Lambda}}\int_{t}^{T}e^{i(8\gamma\zeta^4-2\zeta^2) (t-\tau)\hat{\Lambda}}(R_1G_2)(0,\tau,\Lambda)d\tau,\label{2.19b}\\
&G_3(x,t,\zeta)=\mathrm{I}-\int_{x}^{\infty}e^{-i\zeta(x-\xi)\hat{\Lambda}}(PG_3)(\xi,t,\zeta)d\xi.\label{3.19c}
\end{align}
\end{subequations}

On the one hand, any point $(x,t)$ on the integral curve $\{\gamma_j\}_1^3$ satisfies the following inequalities
\begin{subequations}
\begin{align}
&\gamma_1: x-\xi\geq 0, t-\tau\geq 0,\label{2.20a}\\
&\gamma_2: x-\xi\geq 0, t-\tau\leq 0,\label{2.20b}\\
&\gamma_3: x-\xi\leq 0.\label{2.20c}
\end{align}
\end{subequations}
On the other hand, it follows from the Eq.\eqref{2.18} that the first column of $G_j(x,t,\zeta)$ contains $e^{2i\zeta(x-\xi)-2i(8\gamma\zeta^4-2\zeta^2) (t-\tau)}$. Thus, for $\zeta\in \mathrm{C}$, we can calculate the bounded analytic region of $[G_j(x,t,\zeta)]_1$, that is $\zeta$ must satisfies
\begin{subequations}
\begin{align}
&{[G_{1}]}_1(x,t,\zeta): \{\mathrm{Im}\zeta \geq 0\}\cap\{\mathrm{Im}(8\gamma\zeta^{4}-2\zeta^2)\geq 0\},\label{2.21a}\\
&{[G_{2}]}_1(x,t,\zeta): \{\mathrm{Im}\zeta \geq 0\}\cap\{\mathrm{Im}(8\gamma\zeta^{4}-2\zeta^2)\leq 0\},\label{2.21b}\\
&{[G_{3}]}_1(x,t,\zeta): \{\mathrm{Im}\zeta \leq 0\}.\label{2.21c}
\end{align}
\end{subequations}
Similarly, it follows from the Eq.\eqref{2.18} that the second column of $G_j(x,t,\zeta)$ contains $e^{-2i\zeta(x-\xi)+2i(8\gamma\zeta^4-2\zeta^2) (t-\tau)}$. Then, for $\zeta\in \mathrm{C}$, we can also calculate the bounded analytic region of the eigenfunctions $[G_j(x,t,\zeta)]_2$, that means $\zeta$ must satisfies
\begin{subequations}
\begin{align}
&{[G_{1}]}_2(x,t,\zeta): \{\mathrm{Im}\zeta \leq 0\}\cap\{\mathrm{Im}(8\gamma\zeta^{4}-2\zeta^2)\leq 0\},\label{2.22a}\\
&{[G_{2}]}_2(x,t,\zeta): \{\mathrm{Im}\zeta \leq 0\}\cap\{\mathrm{Im}(8\gamma\zeta^{4}-2\zeta^2)\geq 0\},\label{2.22b}\\
&{[G_{3}]}_2(x,t,\zeta): \{\mathrm{Im}\zeta \geq 0\}.\label{2.22c}
\end{align}
\end{subequations}
where the $[G_j]_{k}(x,t,\zeta)$ denotes the $k$-columns of $G_j(x,t,\zeta)$. After calculation, we get the bounded analytic region of $G_j(x,t,\zeta)$ as follows
\begin{subequations}
\begin{align}
&G_1(x,t,\zeta)=([G_{1}]_1^{L_1\cup L_3}(x,t,\zeta),[G_{1}]_2^{L_6\cup L_8}(x,t,\zeta)),\label{2.23a}\\
&G_2(x,t,\zeta)=([G_{2}]_1^{L_2\cup L_4}(x,t,\zeta),[G_{2}]_2^{L_5\cup L_7}(x,t,\zeta)),\label{2.23b}\\
&G_3(x,t,\zeta)=([G_{3}]_1^{L_5\cup L_6\cup L_7\cup L_8}(x,t,\zeta),[G_{3}]_2^{L_1\cup L_2\cup L_3\cup L_4}(x,t,\zeta)),\label{2.23c}
\end{align}
\end{subequations}
where $G_{j}^{L_i}(x,t,\zeta)$ represents the bounded analytic region of $\{G_j(x,t,\zeta)\}_1^3$ is $\zeta \in L_i, i=1,2,\ldots,8$, and $L_i, i=1,2,\ldots,8$ are shown in Figure 2.

\begin{figure}
  \centering
  \includegraphics[width=1.8in,height=1.3in]{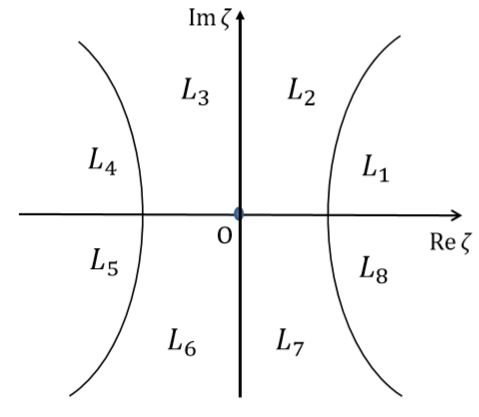}\\
  \caption{The areas $L_i,i=1,\ldots,8$ division on the complex $\zeta$ plane}\label{fig:graph}
\end{figure}

To establish the RH problem of the FODNLS equation \eqref{1.2}, we must also define two special functions $\psi(\zeta)$ and $\phi(\zeta)$ with the eigenfunction $\{G_j(x,t,\zeta)\}_1^3$ as follows
\begin{subequations}
\begin{align}
&G_3(x,t,\zeta)=G_1(x,t,\zeta)e^{i[(8\gamma\zeta^4-2\zeta^2) t-\zeta x]\hat{\Lambda}}\psi(\zeta),\label{2.24a}\\
&G_2(x,t,\zeta)=G_1(x,t,\zeta)e^{i[(8\gamma\zeta^4-2\zeta^2) t-\zeta x]\hat{\Lambda}}\phi(\zeta).\label{2.24b}
\end{align}
\end{subequations}
Set $(x,t)=(0,0)$ in Eq.\eqref{2.24a}, and let $(x,t)=(0,T)$ in Eq.\eqref{2.24b}, we obtain the following relationship
\eqa \psi(\zeta)=G_3(0,0,\zeta), \,\,
\phi(\zeta)=G_2(0,0,\zeta)=[e^{i(8\gamma\zeta^4-2\zeta^2) T\hat{\Lambda}}G_1(0,T,\zeta)]^{-1},
\label{2.25}\eeqa
then, we get
\eqa G_3(x,t,\zeta)=G_1(x,t,\zeta)e^{i[(8\gamma\zeta^4-2\zeta^2) t-\zeta x]\hat{\Lambda}}G_3(0,0,\zeta),\label{2.24c}\eeqa
and
\eqa G_2(x,t,\zeta)=G_1(x,t,\zeta)e^{i[(8\gamma\zeta^4-2\zeta^2) t-\zeta x]\hat{\Lambda}}[e^{-i(8\gamma\zeta^4-2\zeta^2) T\hat{\Lambda}}G_1(0,T,\zeta)]^{-1},\label{2.24d}\eeqa
it follows from the Eqs.\eqref{2.24c}-\eqref{2.24d} that
\eqa G_2(x,t,\zeta)=G_3(x,t,\zeta)e^{i[(8\gamma\zeta^4-2\zeta^2) t-\zeta x]\hat{\Lambda}}(\psi(\zeta))^{-1}\phi(\zeta). \label{2.26}\eeqa
%Furthermore, as a result of $G_2(0,T,\zeta)=\mathrm{I}$, we have the following so-called global relation
%\beq \phi^{-1}(\zeta)\psi(\zeta)=e^{i(8\gamma\zeta^4-2\zeta^2) T\hat{\Lambda}}E(T,\zeta),\label{2.27}\eeq
%where $ E(T,\zeta)=G_3(0,T,\zeta).$

Particularly, in the eigenfunction $G_j(x,t,\zeta),j=1,2$, when $x=0$, we have
\begin{subequations}
\begin{align}
&G_1(0,t,\zeta)=([G_1]_1^{L_1\cup L_3\cup L_5\cup L_7}(0,t,\zeta),[G_1]_2^{L_2\cup L_4\cup L_6\cup L_8}(0,t,\zeta))\nn\\
&\qquad\qquad\;=\mathrm{I}+\int_{0}^{t}e^{i(8\gamma\zeta^4-2\zeta^2) (t-\tau)\hat{\Lambda}}(RG_1)(0,\tau,\zeta)d\tau,\label{2.28a}\\
&G_2(0,t,\zeta)=([G_2]_1^{L_2\cup L_4\cup L_6\cup L_8}(0,t,\zeta),[G_2]_2^{L_1\cup L_3\cup L_5\cup L_7}(0,t,\zeta))\nn\\
&\qquad\qquad\;=\mathrm{I}-\int_{t}^{T}e^{i(8\gamma\zeta^4-2\zeta^2) (t-\tau)\hat{\Lambda}}(RG_2)(0,\tau,\zeta)d\tau,\label{2.28b}
\end{align}
\end{subequations}
and in the eigenfunction $G_1(x,t,\zeta), G_3(x,t,\zeta)$, when $t=0$, we have
\begin{subequations}
\begin{align}
&G_1(x,0;\zeta)=([G_1]_1^{L_1\cup L_2\cup L_3\cup L_4}(x,0,\zeta),[G_1]_2^{L_5\cup L_6\cup L_7\cup L_8}(x,0,\zeta))\nn\\
&\qquad\qquad\;=\mathrm{I}+\int_{0}^{x}e^{-i\zeta(x-\xi)\hat{\Lambda}}(PG_1)(\xi,0,\zeta)d\xi,\label{2.28c}\\
&G_3(x,0;\zeta)=([G_{3}]_1^{L_5\cup L_6\cup L_7\cup L_8}(z,0,\zeta),[G_{3}]_2^{L_1\cup L_2\cup L_3\cup L_4}(z,0,\zeta))\nn\\
&\qquad\qquad\;=\mathrm{I}-\int_{x}^{\infty}e^{-i\zeta(x-\xi)\hat{\Lambda}}(PG_3)(\xi,0,\zeta)d\xi,\label{2.28d}
\end{align}
\end{subequations}
Assuming that $u_0(x)=u(x,t=0)$ is an initial data of the functions $u(x,t)$, and $v_0(t)=u(x=0,t)$, $v_1(t)=u_x(x=0,t)$, $v_2(t)=u_{xx}(x=0,t)$, $v_3(t)=u_{xxx}(x=0,t)$ are boundary datas of the functions $u_x(x,t)$, $u_{xx}(x,t)$, $u_{xxx}(x,t)$, at this time, the matrix $P(x,0,\zeta)$ and $R(0,t,\zeta)$ have the following matrix forms, respectively.
\eqa
P(x,0,\zeta)=\left(\begin{array}{cc}
0 & u_0 \\
-\overline{u}_0  & 0
\end{array}\right),\,\,
R(0,t,\zeta)=\left(\begin{array}{cc}
R_{11}(0,t,\zeta) & R_{12}(0,t,\zeta) \\
R_{21}(0,t,\zeta)  &  -R_{11}(0,t,\zeta)
\end{array}\right),\label{2.28}\eeqa
with
\eqa &&R_{11}(0,t,\zeta)=-4i\gamma\zeta^2v^2_0-2\gamma\zeta(v_0\bar{v}_1-v_1\bar{v}_0)
+i\gamma(3|v_0|^4-|v_1|^2+\overline{v_0}v_{2}+v_0\overline{v}_{2})+i|v_0|^2\nn\\&&
R_{12}(0,t,\zeta)=-8\gamma\zeta^3v_0-4i\gamma\zeta^2v_1+2\gamma\zeta v_{2}+2\zeta(2\gamma|v_0|^2+1)v_0
+i\gamma v_{3}+i(6\gamma|v_0|^2+1)v_1\nn\\&&
R_{21}(0,t,\zeta)=-8\gamma\zeta^3\bar{v}_0-4i\gamma\zeta^2\bar{v}_1+2\gamma\zeta \bar{v}_{2}+2\zeta(2\gamma|v_0|^2+1)\bar{v}_0
+i\gamma \bar{v}_{3}+i(6\gamma|v_0|^2+1)\bar{v}_1\nn\eeqa

\subsection{ The other properties of the eigenfunctions}
\begin{prop}
The matrix-value functions $G_j(x,t,\zeta)=([G_j]_{1}(x,t,\zeta),[G_j]_{2}(x,t,\zeta))(j=1,2,3)$ are given in Eq.\eqref{2.18} enjoy analytical properties are:
\begin{itemize}
  \item $\mathrm{det}G_j(x,t,\zeta)=1$;
  \item The $[G_1]_{1}(x,t,\zeta)$ is an analytic function for $\zeta\in L_1\cup L_3$, and the $[G_1]_{2}(x,t,\zeta)$ is also an analytic function for $\zeta\in L_6\cup L_8$;
  \item The $[G_2]_{1}(x,t,\zeta)$ is an analytic function for $\zeta\in L_2\cup L_4$, and the $[G_2]_{2}(x,t,\zeta)$ is also an analytic function for $\zeta\in L_5\cup L_7$;
  \item The $[G_3]_{1}(x,t,\zeta)$ is an analytic function for $\zeta\in L_5\cup L_6\cup L_7\cup L_8$, and the $[G_3]_{2}(x,t,\zeta)$ is also an analytic function for $\zeta\in L_1\cup L_2\cup L_3\cup L_4$;
  \item The $[G_j]_{1}(x,t,\zeta)\rightarrow(1,0)^T$ and $[G_j]_{2}(x,t,\zeta)\rightarrow(0,1)^T$, as $\zeta\rightarrow\infty$.
\end{itemize}
\end{prop}

\begin{prop}
Indeed, $\psi(\zeta),\phi(\zeta)$ are defined in Eqs.\eqref{2.24a}-\eqref{2.24b} or Eq.\eqref{2.25} is expressed as
\begin{subequations}
\begin{align}
&\psi(\zeta)=\mathrm{I}-\int_{0}^{\infty}e^{i\zeta\xi \hat{\Lambda}}(PG_3)(\xi,0,\zeta)d\xi,\label{2.29a}\\
&\phi^{-1}(\zeta)=\mathrm{I}+\int_{0}^{T}e^{i(8\gamma\zeta^4-2\zeta^2)\tau\hat{\Lambda}}(RG_1)(0,\tau,\zeta)d\tau.\label{2.29b}
\end{align}
\end{subequations}
\end{prop}

It follows from the symmetry properties of $P(x,t,\zeta)$ and $R(x,t,\zeta)$ that
\eqa (G_j(x,t,\zeta))_{11}=\overline{(G_j(x,t,\overline{\zeta}))_{11}},\,\,
(G_j(x,t,\zeta))_{21}=-\overline{(G_j(x,t,\overline{\zeta}))_{12}},
\nn\eeqa
then
$$ \psi_{11}(\zeta)=\overline{\psi_{11}(\overline{\zeta})},\,\,
\psi_{21}(\zeta)=-\overline{\psi_{12}(\overline{\zeta})},$$
$$\phi_{11}(\zeta)=\overline{\phi_{11(\overline{\zeta})}},\,\,
\phi_{21}(\zeta)=-\overline{\phi_{12}(\overline{\zeta})},
$$
and assume that the $\psi(\zeta)$ and $\phi(\zeta)$ admits the matrix form as follows
\eqa
\psi(\zeta)=\left(\begin{array}{cc}
\overline{y(\bar{\zeta})} & z(\zeta)\\
-\overline{z(\bar{\zeta})}& y(\zeta)
\end{array}\right),
\phi(\zeta)=\left(\begin{array}{cc}
\overline{Y(\bar{\zeta})} & Z(\zeta)\\
-\overline{Z(\bar{\zeta})}& Y(\zeta)\\
\end{array}\right).
\label{2.30}\eeqa
In terms of the Eq.\eqref{2.25} and Eqs.\eqref{2.29a}-\eqref{2.29b}, we known that the following properties are ture,
\begin{itemize}
  \item $$\left(\begin{array}{c} z(\zeta) \\ y(\zeta) \\ \end{array} \right)=[G_3]_2^{L_1\cup L_2\cup L_3\cup L_4}(0,0,\zeta)=\left(
   \begin{array}{c}
    (G_3)_{12}^{L_1\cup L_2\cup L_3\cup L_4}(0,0,\zeta) \\
    (G_3)_{22}^{L_1\cup L_2\cup L_3\cup L_4}(0,0,\zeta) \\
  \end{array}
 \right), $$
where the vector function $[G_3]_2^{L_1\cup L_2\cup L_3\cup L_4}(x,0,\zeta)$ satisfy the ordinary differential equation as follows
\eqa && \p_x [G_3]_2^{L_1\cup L_2\cup L_3\cup L_4}(x,0,\zeta)+2i\zeta
\left(\begin{array}{cc}
1& 0\\
0& 0\end{array}\right)[G_3]_2^{L_1\cup L_2\cup L_3\cup L_4}(x,0,\zeta)
\nn\\&&=P(x,0,\zeta)[G_3]_2^{L_1\cup L_2\cup L_3\cup L_4}(x,0,\zeta),\,0<x<\infty,\label{2.30a}\eeqa
where $P(x,0,\zeta)$ is given in Eq\eqref{2.28} and
$$\lim_{x \rightarrow\infty}[G_3]_2^{L_1\cup L_2\cup L_3\cup L_4}(x,0,\zeta)=(0,1)^T.$$
 $$
 \left(\begin{array}{c} -e^{-2i(8\gamma\zeta^4-2\zeta^2) T}Z(\zeta) \\ \overline{Y(\bar{\zeta})} \\ \end{array} \right)
 = [G_1]_2^{L_2\cup L_4\cup L_6\cup L_8}(0,T,\zeta)
 =\left(\begin{array}{c}
    (G_1)_{12}^{L_2\cup L_4\cup L_6\cup L_8}(0,t,\zeta) \\
    (G_1)_{22}^{L_2\cup L_4\cup L_6\cup L_8}(0,t,\zeta) \\
  \end{array}
\right),$$
where the vector function $[G_1]_2^{L_2\cup L_4\cup L_6\cup L_8}(0,t,\zeta)$ satisfy the ordinary differential equation as follows
\eqa && \p_t [G_1]_2^{L_2\cup L_4\cup L_6\cup L_8}(0,t,\zeta)-2i(8\gamma\zeta^4-2\zeta^2)
\left(\begin{array}{cc}
1& 0\\
0& 0\end{array}\right)[G_1]_2^{L_2\cup L_4\cup L_6\cup L_8}(0,t,\zeta)
\nn\\&&=R(0,t,\zeta)[G_1]_2^{L_2\cup L_4\cup L_6\cup L_8}(0,t,\zeta),\,0<t<T,\label{2.30b}\eeqa
where $R(0,t,\zeta)$ is given in Eq\eqref{2.28} and
$$[G_1]_2^{L_2\cup L_4\cup L_6\cup L_8}(0,0,\zeta)=(0,1)^T.$$
 \item $$y(-\zeta)=y(\zeta),\,\, z(-\zeta)=-z(\zeta),$$
 $$ Y(-\zeta)=Y(\zeta),\,\, Z(-\zeta)=-Z(\zeta).$$
 \item $$ For\,\, \zeta\in \mathrm{R},\,\, \mathrm{det} \psi(\zeta)=|y(\zeta)|^2+|z(\zeta)|^2=1.$$
       $$ For\,\, \zeta\in \mathrm{C},\,\, \mathrm{det} \phi(\zeta)=Y(\zeta)\overline{Y(\bar{\zeta})}+ Z(\zeta)\overline{Z(\bar{\zeta})}=1,\,\, (\zeta\in \Gamma_m,\, if\, T=\infty).$$
       where curve $\Gamma_m,\,m=1,2,3,4$ are given in Eq.\eqref{2.35}.
 \item $$ y(\zeta)=1+O(\frac{1}{\zeta}),\,\, z(\zeta)=O(\frac{1}{\zeta}), as\, \zeta\rightarrow\infty, $$
 $$ Y(\zeta)=1+O(\frac{1}{\zeta})+O(\frac{e^{-2i(8\gamma\zeta^4-2\zeta^2) T}}{\zeta}),\,\, Z(\zeta)=O(\frac{1}{\zeta})+O(\frac{e^{-2i(8\gamma\zeta^4-2\zeta^2) T}}{\zeta}), as\, \zeta\rightarrow\infty.$$
 \end{itemize}

\subsection{The basic Riemann-Hilbert problem}

In order to facilitate calculation and formula representation, we introduce the symbolic assumptions as follows
\begin{subequations}
\begin{align}
&\omega(x,t,\zeta)=\zeta x-(8\gamma\zeta^4-2\zeta^2) t,\label{2.31a}\\
&\rho(\zeta)=y(\zeta)\overline{Y(\bar{\zeta})}+z(\zeta)\overline{Z(\bar{\zeta})},\label{2.31b}\\
&\kappa(\zeta)=\overline{y(\bar{\zeta})}\overline{Z(\bar{\zeta})}-\overline{z(\bar{\zeta})}\overline{Y(\bar{\zeta})},\label{2.31c}\\
&\delta(\zeta)=\frac{z(\zeta)}{\overline{y(\bar{\zeta})}},
\Delta(\zeta)=-\frac{\overline{Z(\bar{\zeta})}}{y(\zeta)\rho(\zeta)},\label{2.31e}
\end{align}
\end{subequations}
then, we have
\eqa && \overline{Z(\bar{\zeta})}=y(\zeta)\kappa(\zeta)+\overline{z(\bar{\zeta})}\rho(\zeta),\nn\\&& \rho(\zeta)\overline{\rho(\bar{\zeta})}-\kappa(\zeta)\overline{\kappa(\bar{\zeta})}=1,\nn\\&&
\rho(\zeta)=1+O(\frac{1}{\zeta}), \kappa(\zeta)=O(\frac{1}{\zeta})\,\, as \,\, \zeta\rightarrow\infty,\nn\\&&
\rho(-\zeta)=\rho(\zeta),\kappa(-\zeta)=-\kappa(\zeta),\nn\eeqa
and the matrix function $D(x,t,\zeta)$ is defined by
\begin{subequations}
\begin{align}
&D_{+}(x,t,\zeta)=(\frac{[G_1]_1^{L_1\cup L_3}(x,t,\zeta)}{y(\zeta)},[G_3]_2^{L_1\cup L_2\cup L_3\cup L_4}(x,t,\zeta)),
\zeta\in L_1\cup L_3,\label{2.32a}\\
&D_{-}(x,t,\zeta)=(\frac{[G_2]_1^{L_2\cup L_4}(x,t,\zeta)}{\rho(\zeta)},[G_3]_2^{L_1\cup L_2\cup L_3\cup L_4}(x,t,\zeta)),
\zeta\in L_2\cup L_4,\label{2.32b}\\
&D_{+}(x,t,\zeta)=([G_3]_1^{L_5\cup L_6\cup L_7\cup L_8}(x,t,\zeta),\frac{[G_2]_2^{L_5\cup L_7}(x,t,\zeta)}{\overline{\rho(\bar{\zeta})}}),
\zeta\in L_5\cup L_7,\label{2.32c}\\
&D_{-}(x,t,\zeta)=([G_3]_1^{L_5\cup L_6\cup L_7\cup L_8}(x,t,\zeta),\frac{[G_1]_2^{L_6\cup L_8}(x,t,\zeta)}{\overline{y(\bar{\zeta})}}),
\zeta\in L_6\cup L_8.\label{2.32d}
\end{align}
\end{subequations}
Obviously, the above definitions indicates that
\beq  \mathrm{det} D(x,t,\zeta)=1,\,\, D(x,t,\zeta)\rightarrow\mathrm{I}, as\,\, \zeta\rightarrow\infty. \label{2.33}\eeq

\begin{thm}
Let that the matrix function $D(x,t,\zeta)$ is defined by Eqs.\eqref{2.32a}-\eqref{2.32d} and the potential function $u(z,t)\in \mathcal{S}$, then, the matrix function $D(x,t,\zeta)$ admits the jump relation on the curve $\Gamma_m, m=1,\ldots,4$ as follows
\eqa D_{+}(x,t,\zeta)=D_{-}(x,t,\zeta)Q(x,t,\zeta), \zeta \in \Gamma_m, m=1,\ldots,4,\label{2.34}\eeqa
where
\eqa
Q(x,t,\zeta)=\left\{ \begin{array}{l}
Q_1(x,t,\zeta), \qquad\qquad\qquad   \zeta \in \Gamma_1\doteq\{\bar L_1\cup\bar L_3\}\cap\{\bar L_2\cup\bar L_4\}\\
Q_2(x,t,\zeta)=Q_3Q_4^{-1}Q_1, \quad \zeta \in \Gamma_2\doteq\{\bar L_2\cup\bar L_4\}\cap\{\bar L_5\cup\bar L_7\}\\
Q_3(x,t,\zeta), \qquad\qquad\qquad \zeta \in \Gamma_3\doteq\{\bar L_5\cup\bar L_7\}\cap\{\bar L_6\cup\bar L_8\} \\
Q_4(x,t,\zeta), \qquad\qquad\qquad   \zeta \in \Gamma_4\doteq\{\bar L_6\cup\bar L_8\}\cap\{\bar L_1\cup\bar L_3\}
\end{array}\right.\label{2.35}\eeqa
and
\eqa \begin{array}{l}
Q_1(x,t,\zeta)=\left(\begin{array}{cc}
  1 & 0 \\
\Delta(\zeta)e^{2i\omega(\zeta)} & 1
 \end{array}\right),\\
 Q_2(x,t,\zeta)=\left(\begin{array}{cc}
1-(\delta(\zeta)+\overline{\Delta(\bar{\zeta})})(\Delta(\zeta)+\overline{\delta(\bar{\zeta})}) & (\delta(\zeta)+\overline{\Delta(\bar{\zeta})})e^{-2i\omega(\zeta)} \\
(\Delta(\zeta)+\overline{\delta(\bar{\zeta})})e^{2i\omega(\zeta)} & 1
\end{array}\right),\\
Q_3(x,t,\zeta)=\left(\begin{array}{cc}
1 &\overline{\Delta(\bar{\zeta})}e^{-2i\omega(\zeta)}\\
0 & 1
 \end{array} \right),\\
Q_4(x,t,\zeta)=\left(\begin{array}{cc}
1 & -\delta(\zeta)e^{-2i\omega(\zeta)} \\
-\overline{\delta(\bar{\zeta})}e^{2i\omega(\zeta)} & 1+|\delta(\zeta)|^2
\end{array}\right).
\end{array}\nn\eeqa
\end{thm}
\textbf{Proof} In terms of the Eqs.\eqref{2.24a}-\eqref{2.24b} and Eq.\eqref{2.30}, we have
\begin{subequations}
\begin{align}
&\overline{y(\bar{\zeta})}[G_1]_1^{L_1\cup L_3}(x,t,\zeta)
-\overline{z(\bar{\zeta})}e^{2i\omega(\zeta)}[G_1]_2^{L_6\cup L_8}(x,t,\zeta)
=[G_3]_1^{L_5\cup L_6\cup L_7\cup L_8}(x,t,\zeta), \label{2.36a} \\
&z(\zeta)e^{-2i\omega(\zeta)}[G_1]_1^{L_1\cup L_3}(x,t,\zeta)+y(\zeta)[G_1]_2^{L_6\cup L_8}(x,t,\zeta)
=[G_3]_2^{L_1\cup L_2\cup L_3\cup L_4}(x,t,\zeta),\label{2.36b}
\end{align}
\end{subequations}
and
\begin{subequations}
\begin{align}
&\overline{Y(\bar{\zeta})}[G_1]_1^{L_1\cup L_3}(x,t,\zeta)
-\overline{Z(\bar{\zeta})}e^{2i\omega(\zeta)}[G_1]_2^{L_6\cup L_8}(x,t,\zeta)
=[G_2]_1^{L_2\cup L_4}(x,t,\zeta),\label{2.37a}  \\
&Z(\zeta)e^{-2i\omega(\zeta)}[G_1]_1^{L_1\cup L_3}(x,t,\zeta)+Y(\zeta)[G_1]_2^{L_6\cup L_8}(x,t,\zeta)
=[G_2]_2^{L_5\cup L_7}(x,t,\zeta),\label{2.37b}
\end{align}
\end{subequations}
according to the Eqs.\eqref{2.36a}-\eqref{2.37b} and Eqs.\eqref{2.31a}-\eqref{2.31e} yields
\begin{subequations}
\begin{align}
&\rho(\zeta)[G_3]_1^{L_5\cup L_6\cup L_7\cup L_8}(x,t,\zeta)
-\kappa(\zeta)e^{2i\omega(\zeta)}[G_3]_2^{L_1\cup L_2\cup L_3\cup L_4}(x,t,\zeta)
=[G_2]_1^{L_2\cup L_4}(x,t,\zeta),\label{2.38a}\\
&\overline{\kappa(\bar{\zeta})}e^{-2i\omega(\zeta)}[G_3]_1^{L_5\cup L_6\cup L_7\cup L_8}(x,t,\zeta)
+\overline{\rho(\bar{\zeta})}[G_3]_2^{L_1\cup L_2\cup L_3\cup L_4}(x,t,\zeta)=[G_2]_2^{ L_5\cup L_7}(x,t,\zeta).\label{2.38b}
\end{align}
\end{subequations}
By the Eqs.\eqref{2.32a}-\eqref{2.32d} and Eq.\eqref{2.34}, one have
\begin{subequations}
\begin{align}
&(\frac{[G_1]_1^{L_1\cup L_3}(x,t,\zeta)}{y(\zeta)},[G_3]_2^{L_1\cup L_2\cup L_3\cup L_4}(x,t,\zeta))
=(\frac{[G_2]_1^{L_2\cup L_4}(x,t,\zeta)}{\rho(\zeta)},
[G_3]_2^{L_1\cup L_2\cup L_3\cup L_4}(x,t,\zeta))Q_1(x,t,\zeta),\label{2.39a}\\
&([G_3]_1^{L_5\cup L_6\cup L_7\cup L_8}(x,t,\zeta),\frac{[G_2]_2^{L_5\cup L_7}(x,t,\zeta)}{\overline{\rho(\bar{\zeta})}})
=(\frac{[G_2]_1^{L_2\cup L_4}(x,t,\zeta)}{\rho(\zeta)},
[G_3]_2^{L_1\cup L_2\cup L_3\cup L_4}(x,t,\zeta))Q_2(x,t,\zeta),\label{2.39b}\\
&([G_3]_1^{L_5\cup L_6\cup L_7\cup L_8}(x,t,\zeta),\frac{[G_2]_2^{L_5\cup L_7}(x,t,\zeta)}{\overline{\rho(\bar{\zeta})}})
=([G_3]_1^{L_5\cup L_6\cup L_7\cup L_8}(x,t,\zeta),\frac{[G_1]_2^{L_6\cup L_8}(x,t,\zeta)}{\overline{y(\bar{\zeta})}})Q_3(x,t,\zeta),\label{2.39c}\\
&(\frac{[G_1]_1^{L_1\cup L_3}(x,t,\zeta)}{y(\zeta)},[G_3]_2^{L_1\cup L_2\cup L_3\cup L_4}(x,t,\zeta))
=([G_3]_1^{L_5\cup L_6\cup L_7\cup L_8}(x,t,\zeta),
\frac{[G_1]_2^{L_6\cup L_8}(x,t,\zeta)}{\overline{y(\bar{\zeta})}})Q_4(x,t,\zeta).\label{2.39d}
\end{align}
\end{subequations}
Therefore, we can derive from the Eqs.\eqref{2.39a}-\eqref{2.39d} that the jump matrices $\{Q_i(x,t,\zeta)\}_1^4$ meets the Eq.\eqref{2.35}.
\begin{ass}
Assuming that the zeros of $\rho(\zeta)$ and $y(\zeta)$ enjoy the assumptions as follows
\begin{itemize}
  \item The spectral function $y(\zeta)$ enjoy $2n$ simple zeros $\{\varsigma_j\}_{j=1}^{2n}$, $2n=2n_1+2n_2$, if $\{\varsigma_j\}_1^{2n_1} \in L_1\cup L_3$, then $\{\bar{\varsigma}_j\}_1^{2n_2} \in L_8\cup L_6$.
  \item The spectral function $\rho(\zeta)$ enjoy $2N$ simple zeros $\{\eta_j\}_{j=1}^{2N}$, $2N=2N_1+2N_2$, if $\{\eta_j\}_1^{2N_1} \in L_5\cup L_7$, then $\{\bar{\eta}_j\}_1^{2N_2} \in L_4\cup L_2$.
  \item The spectral function $y(\zeta)$ and the spectral function $\rho(\zeta)$ do not enjoy the same simple zeros.
\end{itemize}
\end{ass}

\begin{prop}
The matrix function $D(x,t,\zeta)$ is defined by Eqs.\eqref{2.32a}-\eqref{2.32d} meets the following residue conditions:
\begin{subequations}
\begin{align}
& \mathrm{Res} \{[D(x,t,\zeta)]_{1} , \varsigma_j\}
=\frac{1}{z(\varsigma_j)\dot{y}(\varsigma_j)}e^{2i\omega(\varsigma_j)}[D(x,t,\varsigma_j)]_{2}, j=1,\cdots,2n_1.\label{2.40a}\\
& \mathrm{Res} \{[D(x,t,\zeta)]_{2} ,  \bar{\varsigma}_j\}
=-\frac{1}{\overline{z(\varsigma_j)}\overline{\dot{y}(\varsigma_j)}}e^{-2i\omega(\bar{\varsigma_j})}[D(x,t,\bar{\varsigma}_j)]_{1}, j=1,\cdots,2n_2.\label{2.40b}\\
& \mathrm{Res} \{[D(x,t,\zeta)]_{1} , \eta_j \}
=-\frac{\overline{Z(\bar{\eta}_j)}}{y(\eta_j)\dot{\rho}(\eta_j)}e^{2i\omega(\eta_j)}[D(x,t,\eta_j)]_{1}, j=1,\cdots,2N_1.\label{2.40c}\\
& \mathrm{Res} \{[D(x,t,\zeta)]_{2} , \ \bar{\eta}_j\}=\frac{Z(\bar{\eta}_j)}{\overline{y(\eta_j)}\overline{\dot{\rho}(\eta_j)}}e^{-2i\omega(\bar{\eta}_j)}[D(x,t,\bar{\eta_j})]_{2}, j=1,\cdots,2N_2.\label{2.40d}
\end{align}
\end{subequations}
where $\dot{\rho}(\zeta)=\frac{d\rho}{d\zeta}$.
\end{prop}

\textbf{Proof} We only manifest that the residue relationship Eq.\eqref{2.40a} as follows:

Due to $D(x,t,\zeta)=(\frac{[G_1]_1^{L_1\cup L_3}}{y(\zeta)},[G_3]_2^{L_1\cup L_2\cup L_3\cup L_4})$, which is means that the zeros $\{\varsigma_j\}_1^{2n_1}$ of $y(\zeta)$ are the poles of $\frac{[G_1]_1^{L_1\cup L_3}}{y(\zeta)}$. Then, we have
\beq \mathrm{Res}\{\frac{G_1^{L_1\cup L_3}(x,t,\zeta)}{y(\zeta)},\varsigma_j\}
=\lim_{\zeta\rightarrow\varsigma_j}(\zeta-\varsigma_j)\frac{[G_1]_1^{L_1\cup L_3}(x,t,\zeta)}{y(\zeta)}
=\frac{[G_1]_1^{L_1\cup L_3}(x,t,\varsigma_j)}{\dot{y}(\varsigma_j)}.\label{2.41}\eeq
Taking $\zeta=\varsigma_j$ into the second equation of Eqs.\eqref{2.38a}-\eqref{2.38b} yields
\beq [G_1]_1^{L_1\cup L_3}(x,t,\varsigma_j)
=\frac{1}{y(\varsigma_j)}e^{2i\omega(\varepsilon_j)}[G_3]_2^{L_1\cup L_2\cup L_3\cup L_4}(x,t,\varsigma_j).\label{2.42}\eeq
According to the Eq.\eqref{2.41} and Eq.\eqref{2.42}, we get
\beq \mathrm{Res}\{\frac{[G_1]_1^{L_1\cup L_3}(x,t,\zeta)}{y(\zeta)},\varsigma_j\}
=\frac{1}{z(\varsigma_j)\dot{y}(\varsigma_j)}e^{2i\omega(\varsigma_j)}[G_3]_2^{L_1\cup L_2\cup L_3\cup L_4}(x,t,\varsigma_j).\label{2.43}\eeq
Therefore, the Eq.\eqref{2.43} can lead to the Eq.\eqref{2.40a}, and the remaining three residue relationships Eqs.\eqref{2.40b}-\eqref{2.40d} can be similarly proved.

\subsection{The global relation  }

In this subsection, we give the spectral functions are not independent but meet a nice global relation. In fact, at the boundary of the region ${(\xi,\tau): 0<\xi<\infty, 0<\tau<t}$, the integral of the one-form $F(x,t,\zeta)$ is given by the Eq.\eqref{2.7} is vanished. If we assume $G(x,t,\zeta)=G_3(x,t,\zeta)$ in the one-form $F(x,t,\zeta)$ is given by the Eq.\eqref{2.7}, one can get
\eqa && \int_{\infty}^{0}e^{i\zeta\xi\hat{\Lambda}}(PG_3)(\xi,0,\zeta)d\xi
+\int_{0}^{t}e^{-i(8\gamma\zeta^4-2\zeta^2)\tau\hat{\Lambda}}(RG_3)(0,\tau,\zeta)d\tau
+e^{-i(8\gamma\zeta^4-2\zeta^2) t\hat{\Lambda}}\times\int_{0}^{\infty}e^{i\zeta\xi\hat{\Lambda}}(PG_3)(\xi,t,\zeta)d\xi\nn\\&&
=\lim_{x\rightarrow\infty}e^{i\zeta x\hat{\Lambda}}\int_{0}^{t}e^{-i(8\gamma\zeta^4-2\zeta^2)\tau\hat{\Lambda}}(RG_3)(x,\tau,\zeta)d\tau.
\label{2.44}\eeqa
On the one hand, according to the definition of $\psi(\zeta)$ in Eq.\eqref{2.25} and together with the Eq.\eqref{2.28d}, we known that the first term of the Eq.\eqref{2.44} is
$$\psi(\zeta)-I.$$
Let $x=0$ in the Eq.\eqref{2.24c} to get
\eqa G_3(0,\tau,\zeta)=G_1(0,\tau,\zeta)e^{i(8\gamma\zeta^4-2\zeta^2)\tau\hat{\Lambda}}\psi(\zeta),  \label{2.45}\eeqa
therefore
\eqa e^{-i(8\gamma\zeta^4-2\zeta^2)\tau\hat{\Lambda}}(RG_3)(0,\tau,\zeta)=[e^{-i(8\gamma\zeta^4-2\zeta^2)\tau\hat{\Lambda}}(RG_1)(0,\tau,\zeta)]\psi(\zeta).  \label{2.46}\eeqa
On the other hand, the Eq.\eqref{2.46} and Eq.\eqref{2.28a} means that the second term of the Eq.\eqref{2.44} is
\eqa \int_{0}^{t}e^{-i(8\gamma\zeta^4-2\zeta^2)\tau\hat{\Lambda}}(RG_3)(0,\tau,\zeta)d\tau
=[e^{-i(8\gamma\zeta^4-2\zeta^2) t\hat{\Lambda}}RG_1(0,t,\zeta)-I]\psi(\zeta).  \nn\eeqa
For $x\rightarrow\infty$, setting $u(x,t)\in \mathcal{S}$, then, the Eq.\eqref{2.44} is equivalent to
\eqa \phi^{-1}(t,\zeta)\psi(\zeta)+e^{-i(8\gamma\zeta^4-2\zeta^2) t\hat{\Lambda}}\times
\int_{0}^{\infty}e^{i\zeta\xi\hat{\Lambda}}(PG_3)(\xi,t,\zeta)d\xi=I,\label{2.48}\eeqa
where the first column of the Eq.\eqref{2.48} is valid for $\zeta\in L_5\cup L_6\cup L_7\cup L_8$ and the second column of the Eq.\eqref{2.48} is valid for $\zeta\in L_1\cup L_2\cup L_3\cup L_4$, and $\phi(t,\zeta)$ is given by
\eqa \phi^{-1}(t,\zeta)=e^{-i(8\gamma\zeta^4-2\zeta^2) t\hat{\Lambda}}G_1(0,t,\zeta),\nn\eeqa
Owing to $\phi(\zeta)=\phi(T,\zeta)$ and letting $t=T$, then, the Eq.\eqref{2.48} is equivalent to
\eqa \phi^{-1}(\zeta)\psi(\zeta)+e^{-i(8\gamma\zeta^4-2\zeta^2) T\hat{\Lambda}}\times
\int_{0}^{\infty}e^{i\zeta\xi\hat{\Lambda}}(PG_3)(\xi,T,\zeta)d\xi=I.\label{2.49}\eeqa
Hence, the (12)th-component of the Eq.\eqref{2.49} equals
\eqa y(\zeta)Z(\zeta)-Y(\zeta)z(\zeta)=e^{-2i(8\gamma\zeta^4-2\zeta^2) T}E(\zeta),\label{2.50}\eeqa
where $E(\zeta)$ expressed as
\eqa E(\zeta)=\int_{0}^{\infty}e^{i\zeta\xi}(PG_3)_{12}(\xi,T,\zeta)d\xi,\label{2.51}\eeqa
which is the so-called global relation.

\section{The spectral functions }

\begin{de}
 (Related to $y(\zeta)$, $z(\zeta)$) Let $u_{0}(x)=u(x,0)\in \mathcal{S}$, the map
$$\mathrm{H_1}: \{u_0(x)\}\rightarrow \{y(\zeta),z(\zeta) \},$$
is defined by
$$
\left(\begin{array}{c}
z(\zeta) \\
y(\zeta) \end{array}\right)
=[G_3]_2^{L_1\cup L_2\cup L_3\cup L_4}(0;\zeta)=\left(\begin{array}{c}
(G_3)_{12}^{L_1\cup L_2\cup L_3\cup L_4}(0;\zeta) \\
(G_3)_{22}^{L_1\cup L_2\cup L_3\cup L_4}(0;\zeta) \end{array}\right),  \mathrm{Im}\zeta \geq 0,
$$
where $G_3(x,\zeta)$ with $P(x,\zeta)$ are given by Eq.\eqref{2.28d} and Eq.\eqref{2.28}, respectively.
\end{de}

\begin{prop}
The $y(\zeta)$ and $z(\zeta)$ satisfies the properties as follows
\begin{description}
  \item[(i)] For $\mathrm{Im}\zeta <0$, $y(\zeta)$ and $z(\zeta)$ are analytic functions,
  \item[(ii)] $y(\zeta)=1+O(\frac{1}{\zeta}), z(\zeta)=O(\frac{1}{\zeta}),$ as $ \zeta\rightarrow\infty$,
  \item[(iii)] For $\zeta\in \mathrm{R},\,\, \mathrm{det} \psi(\zeta)=|y(\zeta)|^2+|z(\zeta)|^2=1$,
  \item[(iv)] $\mathrm{S_1}=\mathrm{H_1}^{-1}: \{y(\zeta),z(\zeta) \}\rightarrow \{u_0(x)\}$, it's defined as follows
$$u_0(x)=2i\lim_{\zeta\rightarrow\infty}(\zeta D^{(x)}(x,\zeta))_{12},$$
 where $D^{(x)}(x,\zeta)$ meet the following RH problem.
\end{description}
\end{prop}

\begin{rem}
Assume that
\begin{subequations}
\begin{align}
&D_{+}^{(x)}(x,\zeta)=(\frac{[G_{1}]_1^{L_1\cup L_2\cup L_3\cup L_4}(x,\zeta)}{y(\zeta)},
[G_{3}]_2^{L_1\cup L_2\cup L_3\cup L_4}(x,\zeta)),\,\mathrm{Im}\zeta\geq 0,\label{3.2a}\\
&D_{-}^{(x)}(x,\zeta)=([G_{3}]_1^{L_5\cup L_6\cup L_7\cup L_8}(x,\zeta),\frac{[G_{1}]_2^{L_5\cup L_6\cup L_7\cup L_8}(x,\zeta)}{\overline{y(\bar{\zeta})}}),\,\mathrm{Im}\zeta\leq 0,\label{3.2b}
\end{align}
\end{subequations}
hence, $D^{(x)}(x,\zeta)$ admits the RH problem as:
\end{rem}

\begin{itemize}
  \item $D^{(x)}(x,\zeta)=\left\{ \begin{array}{l}
  D_{+}^{(x)} (x,\zeta),  \zeta\in L_1\cup L_2\cup L_3\cup L_4 \\
  D_{-}^{(x)} (x,\zeta),  \zeta\in L_5\cup L_6\cup L_7\cup L_8
\end{array}\right.$ is a slice analytic function.
  \item $D_{+}^{(x)}(x,\zeta)=D_{-}^{(x)}(x,\zeta)(Q^{(x)} (x,\zeta))^{-1}$, $\zeta\in \mathrm{R}$, and
\eqa
Q^{(x)} (x,\zeta)=\left(\begin{array}{cc}
\frac{1}{y(\zeta)\overline{y(\bar{\zeta})}} & \frac{z(\zeta)}{\overline{y(\bar{\zeta})}}e^{-2i\zeta x} \\
-\frac{\overline{z(\bar{\zeta})}}{y(\zeta)}e^{2i\zeta x} & 1
\end{array}\right). \label{3.3}\eeqa
  \item $D^{(x)} (x,\zeta)\rightarrow\mathrm{I}, \zeta\rightarrow\infty.$
  \item $y(\zeta)$ possess $2n$ simple zeros $\{\varsigma_j\}_{1}^{2n}$, $2n=2n_1+2n_2$, let us pretend that $\{\varsigma_j\}_1^{2n_1}$  be part of $L_1\cup L_2\cup L_3\cup L_4$, then, $\{\bar{\varsigma}_j\}_1^{2n_2}$ be part of $L_5\cup L_6\cup L_7\cup L_8$.
  \item $[D_{+}^{(x)}]_1(x,\zeta)$ enjoy simple poles for $\zeta=\{\varsigma_j\}_1^{2n_1}$ and the $[D_{-}^{(x)}]_1(x,\zeta)$ enjoy simple poles for $\zeta=\{\bar{\varsigma}_j\}_1^{2n_2}$. In this case, the residue relations define by
\begin{subequations}
\begin{align}
&\mathrm{Res} \{[D^{(x)}(x;\zeta)]_{1} , \varsigma_j\}
=\frac{e^{2i\varsigma_jx}}{z(\varsigma_j)\dot{y}(\varsigma_j)}[D^{(x)}(x;\varsigma_j)]_{2},\label{3.4a}\\
&\mathrm{Res} \{[D^{(x)}(x;\Lambda)]_{2} ,  \bar{\varsigma}_j\}
=\frac{e^{-2i\bar{\varsigma}_jx}}{\overline{z(\varsigma_j)}\overline{\dot{y}(\varsigma_j)}}[D^{(y)}(x;\bar{\varsigma}_j)]_{1}.\label{3.4b}
\end{align}
\end{subequations}
\end{itemize}

\begin{de}
(Related to  $Y(\zeta)$, $Z(\zeta)$). Set $v_{0}(t)$, $v_{1}(t),v_2(t),v_3(t)\in \mathcal{S}$, the map
$$\mathrm{H_2}: \{v_0(t),v_1(t),v_2(t),v_3(t)\}\rightarrow \{Y(\zeta),Z(\zeta) \},$$
is defined by
$$\left(\begin{array}{c}
Y(\zeta) \\
Z(\zeta) \end{array}\right)
={[G_2]_2}^{L_2\cup L_4\cup L_6\cup L_8}(t,\zeta)=\left(\begin{array}{c}
(G_2)_{12}^{L_2\cup L_4\cup L_6\cup L_8}(t,\zeta) \\
(G_2)_{22}^{L_2\cup L_4\cup L_6\cup L_8}(t,\zeta)\end{array}\right),  $$
where $G_2(t,\zeta)$ with $R(t,\zeta)$ are given by Eq.\eqref{2.28b} and Eq.\eqref{2.28}, respectively.
\end{de}

\begin{prop}
The $Y(\zeta)$ and $Z(\zeta)$ satisfies the properties as follows
\begin{description}
  \item[(i)] For $\mathrm{Im}(8\gamma\zeta^4-2\zeta^2) \leq 0$, $Y(\zeta), Z(\zeta)$ are analytic functions,
  \item[(ii)] $Y(\zeta)=1+O(\frac{1}{\zeta})+O(\frac{e^{-2i(8\gamma\zeta^4-2\zeta^2) T}}{\zeta}), Z(\zeta)=O(\frac{1}{\zeta})+O(\frac{e^{-2i(8\gamma\zeta^4-2\zeta^2) T}}{\zeta}),$ as $ \zeta\rightarrow\infty$,
  \item[(iii)] For $\zeta\in \mathrm{C},\,\, \mathrm{det}\phi(\zeta)=Y(\zeta)\overline{Y(\bar{\zeta})}
      +Z(\zeta)\overline{Z(\bar{\zeta})}=1,\,\,((8\gamma\zeta^4-2\zeta^2)\in \mathrm{R},\, if\, T=\infty)$,
  \item[(iv)]  $\mathrm{S_2}=\mathrm{H_2}^{-1}: \{Y(\zeta),Z(\zeta) \}\rightarrow \{v_0(t),v_1(t),v_2(t),v_3(t)\}$, is defined by
\begin{subequations}
\begin{align}
& v_0(t)=2i(D^{(1)}(t))_{12}=2i\lim_{\zeta\rightarrow\infty}(\zeta D^{(t)}(t,\zeta))_{12},\label{3.5a}\\
& v_1(t)=4(D^{(2)}(t))_{12}+2iv_0(t)(D^{(1)}(t))_{12}
=\lim_{\zeta\rightarrow\infty}[4(\zeta^2 D^{(t)}(t,\zeta))_{12}+2iv_0(t)(\zeta D^{(t)}(t,\zeta))_{22}],\label{3.5b}\\
& \gamma v_2(t)=-8i\gamma(D^{(3)}(t))_{12}+4\gamma v_0(t)(D^{(2)}(t))_{22}
+2i\gamma(v_0^2(t)(D^{(1)}(t))_{12}+v_1(t)(D^{(1)}(t))_{22})\nn\\&\qquad\quad
+2i(D^{(1)}(t))_{12}-(2\gamma v_0^2(t)+1)v_0(t)\nn\\&\qquad
=\lim_{\zeta\rightarrow\infty}[-8i\gamma(\zeta^3D^{(t)}(t,\zeta))_{12}+4\gamma v_0(t)(\zeta^2D^{(t)}(t,\zeta))_{22}
+2i\gamma(v_0^2(t)(\zeta D^{(t)}(t,\zeta))_{12}+v_1(t)(\zeta D^{(t)}(t,\zeta))_{22})\nn\\&\qquad\quad
+2i(\zeta D^{(t)}(t,\zeta))_{12}-(2\gamma v_0^2(t)+1)v_0(t)],\label{3.5c}\\
& \gamma v_3(t)=-16\gamma (D^{(4)}(t))_{12}+4(D^{(2)}(t))_{12}-8i\gamma v_0(t)(D^{(3)}(t))_{22}
+4\gamma(v_0^2(t)(D^{(2)}(t))_{12}+v_1(t)(D^{(2)}(t))_{22})\nn\\&\qquad\quad+i(6\gamma v_0^2(t)+1)v_1(t)
-2i[\gamma(v_0(t)\bar{v_1}(t)-\bar{v_0}(t)v_1(t))(D^{(1)}(t))_{12}-(\gamma v_2(t)
+(2\gamma v_0^2(t)+1)v_0(t))(D^{(1)}(t))_{22}]\nn\\&\qquad
=\lim_{\zeta\rightarrow\infty}\{-16\gamma(\zeta^4D^{(t)}(t,\zeta))_{12}
+4(\zeta^2D^{(t)}(t,\zeta))_{12}-8i\gamma v_0(t)(\zeta^3D^{(t)}(t,\zeta))_{22}\nn\\&\qquad\quad
+4\gamma(v_0^2(t)(\zeta^2D^{(t)}(t,\zeta)_{12}+v_1(t)(\zeta^2D^{(t)}(t,\zeta))_{22})+i(6\gamma v_0^2(t)+1)v_1(t)\nn\\&\qquad\quad
-2i[\gamma(v_0(t)\bar{v_1}(t)-\bar{v_0}(t)v_1(t))(\zeta D^{(t)}(t,\zeta))_{12}-(\gamma v_2(t)
+(2\gamma v_0^2(t)+1)v_0(t))(\zeta D^{(t)}(t,\zeta))_{22}]\},\label{3.5d}
\end{align}
\end{subequations}
where $D^{(1)}(t), D^{(2)}(t), D^{(3)}(t),D^{(4)}(t)$ meets the following asymptotic expansion
$$D^{(t)}(t,\zeta)=\mathrm{I}+\frac{D^{(1)}(t,\zeta)}{\zeta}+\frac{D^{(2)}(t,\zeta)}{\zeta^2}
+\frac{D^{(3)}(t,\zeta)}{\zeta^3}+\frac{D^{(4)}(t,\zeta)}{\zeta^4}+O(\frac{1}{\zeta^5}),\,as\,\zeta\rightarrow\infty,$$
where $D^{(t)}(t,\zeta)$ meet the following RH problem:
\end{description}
\end{prop}

\begin{rem} Assume that
\begin{subequations}
\begin{align}
&D_{+}^{(t)}(t,\zeta)=(\frac{[G_{1}]_1^{L_1\cup L_3\cup L_5\cup L_7}(t,\zeta)}{Y(\zeta)},
[G_{3}]_2^{L_1\cup L_3}(t,\zeta)),\,\mathrm{Im}(8\gamma\zeta^4-2\zeta^2) \geq 0,\label{3.6a}\\
&D_{-}^{(t)}(t,\zeta)=([G_{3}]_1^{L_2\cup L_4}(t,\zeta),\frac{[G_{1}]_2^{L_2\cup L_4}(t,\zeta)}{\overline{Y(\bar{\zeta})}}),\,\mathrm{Im}(8\gamma\zeta^4-2\zeta^2) \leq 0,\label{3.6b}
\end{align}
\end{subequations}
hence, $D^{(t)}(t,\zeta)$ admits the RH problem as follows.
\end{rem}

\begin{itemize}
  \item $D^{(t)}(t,\zeta)=\left\{\begin{array}{ll}
D_{+}^{(t)} (t,\zeta), & \zeta\in L_1\cup L_4\cup L_5 \cup L_8  \\
D_{-}^{(t)} (t,\zeta), & \zeta\in L_2\cup L_3\cup L_6 \cup L_7
\end{array}\right.$ is a sectionally analytic function.
  \item $D_{+}^{(t)} (t,\zeta)=D_{-}^{(t)} (t,\zeta)Q^{(t)} (t,\zeta)$, $\zeta\in \Gamma_n,\,n=1,2,3,4$, and
\eqa
Q^{(t)} (t,\zeta)=\left(\begin{array}{cc}
1 & -\frac{Z(\zeta)}{\overline{Y(\bar{\zeta})}}e^{2i(8\gamma\zeta^4-2\zeta^2) t} \\
-\frac{\overline{Z(\bar{\zeta})}}{Y(\zeta)}e^{-2i(8\gamma\zeta^4-2\zeta^2) t} & \frac{1}{Y(\zeta)\overline{Y(\bar{\zeta})}}
\end{array}\right).\label{3.7}\eeqa
  \item $D^{(t)} (t,\zeta)\rightarrow\mathrm{I}, \zeta\rightarrow\infty.$
  \item $Y(\zeta)$ posses $2k$ simple zeros $\{\nu_j\}_{1}^{2k}$, $2k=2k_1+2k_2$, let us pretend that $\{\nu_j\}_1^{2k_1}$ be part of $L_1\cup L_3$, then, $\{\bar{\nu}_j\}_1^{2k_2}$ be part of $L_2\cup L_4$.
  \item $[D_{+}^{(t)}]_1(t,\zeta)$ enjoy simple poles for $\zeta=\{\nu_j\}_1^{2k_1}$ and the $[D_{-}^{(t)}]_2(t,\Lambda)$ enjoy simple poles for $\zeta=\{\bar{\nu}_j\}_1^{2k_2}$. In this case, the residue relations define by
\begin{subequations}
\begin{align}
&\mathrm{Res} \{[D^{(t)}(t,\zeta)]_{1} , \nu_j\}
=\frac{e^{-2(8\gamma\nu_j^{4}-2\nu_j^2)t}}{Z(\nu_j)\dot{Y}(\nu_j)}[D^{(t)}(t,\nu_j)]_{2},\label{3.8a}\\
&\mathrm{Res} \{[D^{(t)}(t,\zeta)]_{2} ,  \bar{\nu}_j\}
=\frac{e^{2(8\gamma\bar{\nu}_j^{4}-2\bar{\nu}_j^2)t}}{\overline{Z(\bar{\nu}_j)}\overline{\dot{Y}(\nu_j)}}[D^{(t)}(t,\nu_j)]_{1}.\label{3.8b}
\end{align}
\end{subequations}
\end{itemize}

\section{The Riemann-Hilbert problem}
In this part, we give two important results in theorem form.
\begin{thm}
Set $u_0(x)\in \mathcal{S}(\mathrm{R^{+}})$, and $\psi(\zeta),\phi(\zeta)$ is defined in terms of $y(\zeta)$, $z(\zeta)$, $Y(\zeta), Z(\zeta)$ are showed in Eq.\eqref{2.30}, respectively. And the $y(\zeta)$, $z(\zeta)$, $Y(\zeta), Z(\zeta)$ are denotes by functions $u_0(z)$, $v_j(t), j=0,\ldots,3$ are showed in \textbf{Definition 3.1} and \textbf{Definition 3.4}. Assume that the function $y(\zeta)$ possess the possible simple zeros are $\{\varsigma_j\}_{j=1}^{2n}$, and the function $\rho(\zeta)$ possess the possible simple zeros are $\{\eta_j\}_{j=1}^{2N}$. Therefore, the solution of the FODNLS equation \eqref{1.2} is
\eqa u(x,t)=2i\lim_{\zeta\rightarrow\infty}(\zeta D(x,t,\zeta))_{12},\label{4.0}\eeqa
where $D(x,t,\zeta)$ is the solution of the RH problems as follows:
\begin{itemize}
  \item $D(x,t,\zeta)$ is a piecewise analytic function for $\zeta\in C\backslash\Gamma_m\,(m=1,\ldots,4)$.
  \item $D(x,t,\zeta)$ jump appears on the curves $\Gamma_m,$ which meets the jump conditions as
\beq D_{+}(x,t,\zeta)=D_{-}(x,t,\zeta)Q(x,t,\zeta),\,\zeta \in \Gamma_m, m=1,\ldots,4\label{4.1}\eeq
\item $D(x,t,\zeta)=\mathrm{I}+O(\frac{1}{\zeta}),\,\zeta\rightarrow\infty$.
\item $D(x,t,\zeta)$ possess residue relationship are showed in \textbf{Proposition 2.5}.
\end{itemize}
Thus, the matrix function $D(x,t,\zeta)$ exists and is unique. Furthermore
$$u(x,0)=u_0(x),\,\,u(0,t)=v_0(t),\,\,u_x(0,t)=v_1(t),\,\,u_{xx}(0,t)=v_2(t),\,\,u_{xxx}(0,t)=v_3(t).$$
\end{thm}

\textbf{Proof.} In fact, the manifest of this RH problem by following\cite{Fokas2005}.

\begin{thm}(The vanishing theorem) If the matrix function $D(x,t,\zeta)\rightarrow 0 \,\, (\zeta \rightarrow \infty),$ then, the RH problem in \textbf{Theorem 4.1} possess only the zero solution.
\end{thm}

\textbf{Proof.} Indeed, the derivation of this vanishing theorem is given in\cite{Fokas2005}.

\begin{rem}
So far, we have obtained the RH problem of Eq.\eqref{1.2} on the half-line, when $\gamma=0$, that is the IBVPs of the standard NLS equation case\cite{Fokas2005}. Different from the standard NLS equation, where the bounded analytical region and the jump curve of the FODNLS equation \eqref{1.2} are different. which the jump curve contains not only the coordinate axis, but also the hyperbola on the $x$-axis, and the analytical region is not symmetrical.
\end{rem}

\section{Conclusions}

In this paper, we utilize Fokas method to investigate integrable FODNLS equation\eqref{1.2}, which can simulate the nonlinear transmission and interaction of ultrashort pulses in the high-speed optical fiber transmission system, and describe the nonlinear spin excitation phenomenon of one-dimensional Heisenberg ferromagnetic chain with eight poles and dipole interaction. Introduce a slice of important functions to spectral analysis of the Lax pair, established the basic RH problem, and the global relationship between spectral functions is also given. Furthermore, we can analyze the integrable FODNLS equation\eqref{1.2} on a finite interval, and also discuss the asymptotic behavior for the solution of the integrable FODNLS equation\eqref{1.2}. These two questions will be studied in our future investigation.

\section*{Acknowledgements}

This work has been partially supported by the NSFC (Nos. 11601055, 11805114 and 11975145), the NSF of Anhui Province (No.1408085QA06), the University Excellent Talent Fund of Anhui Province (No. gxyq2019096), the Natural Science Research Projects in Colleges and Universities of Anhui Province (No. KJ2019A0637).

\end{document}